\documentclass[12pt]{iopart}

\usepackage{amssymb}
\usepackage{graphicx}
\usepackage[colorlinks=true,allcolors=blue]{hyperref}

\newcommand{\ii}{\mathrm{i}}
\newcommand{\jj}{\mathrm{j}}
\newcommand{\kk}{\mathrm{k}}

\begin{document}

\title{Fast reconstruction of single-shot wide-angle diffraction images through 
deep learning}
\author{T Stielow, R Schmidt, C Peltz, T Fennel and S Scheel}
\address{Institut f\"ur Physik, Universit\"at Rostock,
Albert-Einstein-Stra{\ss}e 23--24, D-18059 Rostock, Germany}
\ead{thomas.stielow@uni-rostock.de}
\ead{r.schmidt@digital-ratio.de}
\ead{stefan.scheel@uni-rostock.de}

\begin{abstract}
Single-shot X-ray imaging of short-lived nanostructures such as clusters and 
nanoparticles near a phase transition or non-crystalizing objects such as large 
proteins and viruses is currently the most elegant method for characterizing
their structure. Using hard X-ray radiation provides scattering images that 
encode two-dimensional projections,
which can be combined to identify the full three-dimensional object structure 
from multiple identical samples.
Wide-angle scattering using XUV or soft X-rays, despite yielding lower 
resolution, provides three-dimensional structural information in a single shot 
and has opened routes towards the characterization of non-reproducible objects 
in the gas phase. The retrieval of the structural information contained in 
wide-angle scattering images is highly non-trivial, and currently no efficient 
rigorous
algorithm is known. Here we show that deep learning networks, trained with 
simulated scattering data, allow for fast and accurate reconstruction of
shape and orientation of nanoparticles from experimental images. The gain in 
speed compared to conventional retrieval techniques opens the route for 
automated structure reconstruction algorithms capable of real-time 
discrimination and pre-identification of nanostructures in scattering 
experiments with high repetition rate -- thus representing the enabling 
technology for fast femtosecond nanocrystallography.
\end{abstract}

\submitto{Machine Learning: Science and Technology}

\maketitle

\section{Introduction}
Sources of soft and hard X-rays with large photon flux such as free electron 
lasers \cite{Chapman2006,Gaffney2007} have enabled the high-resolution imaging 
of unsupported nanosystems such as viruses \cite{Seibert2011}, helium 
droplets \cite{Gomez2014,Rupp_2017,Langbehn_2018}, rare-gas 
clusters \cite{Rupp2012}, or metallic nanoparticles \cite{Barke_2015}.
For reproducible samples, a set of scattering images taken for different 
orientations in the small-angle scattering limit, each delivering a 
two-dimensional 
projection of the object's density, can be used to retrieve its 
three-dimensional structure using conventional reconstruction algorithms 
\cite{Ekeberg_2015, Ayyer_2019}. 
Short-lived and non-reproducible objects, however, 
elude the repeated acquisition of several images required for the tomographic 
reconstruction from small-angle scattering. The partial three-dimensional 
information contained in wide-angle scattering enables to overcome this main
deficiency, for the prize of an even more complicated inversion problem
\cite{Rupp_2017,Barke_2015,Raines2010}. Finding a fast reconstruction method 
thus remains the major obstacle for exploiting the potential of wide-angle 
scattering for genuine single-shot structure characterization.

Two aspects distinguish wide-angle from small-angle scattering. First, the 
projection approximation is no longer valid due to substantial contributions 
of the longitudinal component of the wavevector, such that the curvature of the 
Ewald sphere plays an important role. Second, for the wavelength range for 
which wide-angle scattering is realized, the refractive index of most materials 
deviates substantially from unity, and hence multiple scattering, absorption,
backpropagating waves, and refraction all have to be accounted for. Currently, 
all these constraints can only be met by solving the full three-dimensional 
scattering problem by, e.g., finite-difference time-domain (FDTD) methods, 
gridless discrete-dipole approximation (DDA) techniques, or appropriate 
approximate solutions based on multislice Fourier transform (MSFT) 
techniques \cite{Langbehn_2018, Gessner2019}.

These methods allow, for an assumed geometry model of the nanoparticle, to 
describe their wide-angle scattering patterns. However, the determination of 
the geometry from those patterns is highly nontrivial, as there exists no 
rigorous inversion method. Consequently, the existing applications of 
wide-angle scattering had to be based on a parametrized geometry model whose 
parameters can be determined by an iterative forward fit, e.g. by an ensemble of 
optimization trajectories in phase space as employed in the simplex Monte Carlo 
approach in \cite{Langbehn_2018}. Because for every iteration step, at 
least one forward simulation has to be performed, this method is only applicable 
to a small data set and for a sufficiently simple geometry model 
\cite{Langbehn_2018}. Hence, there is an urgent need for efficient 
reconstruction methods that can be used in real time for a large data set.
Here we present a proof-of-principle study that shows, by considering 
icosahedra, that a neural network, trained with simulated scattering images, 
establishes a high-quality reconstruction method of particle size and 
orientation with unprecedented speed.

Machine learning using neural networks, and deep learning in particular, are 
ideally suited for the extraction of structural parameters from scattering 
images, as this is equivalent to the retrieval of a small number of parameters 
or classes from high-dimensional spaces \cite{hinton2006,lecun2015}.
Originally conceived for analyzing big data, deep learning has already had
significant impact in natural sciences, ranging from analyzing phase transitions
and properties of matter \cite{carleo2019review, wang2016,carrasquilla2017,van2017,venderley2018}
and simulations of many-body quantum systems \cite{carleo2017} to quantum state
reconstruction \cite{torlai2018,xin2018}.
In X-ray imaging, neural networks have been introduced only recently. 
In the small-angle regime, it has been demonstrated that both the 
tasks of phase retrieval and reconstruction of theoretically generated
binary two-dimensional density distributions can be solved also with neural networks 
\cite{Cherukara_2018}.
From transmission scattering patterns of thin materials 
lattice cell orientation maps were extracted \cite{laanait2019}, as a parameter
representation of the material structure. In a next step, using the computational power of the world's largest 
supercomputer, full density projection images have been recovered 
\cite{laanait2019exascale}. In addition, the task of pre-classifying scattering 
patterns has been successfully tackled with neural networks 
\cite{Langbehn_2018}, and reinforcement learning techniques have provided 
further insights into experimental features of X-ray scattering patterns 
\cite{Zimmermann2019}. In these previous applications, the neural networks 
were either trained and tested solely on theoretical data or were used for 
feature extraction from experimental diffraction images. In our work 
we take the decisive step by training a neural network on augmented theoretical 
data and use it for predictions on experimental scattering data.


\section{Experimental and theoretical framework}
\begin{figure}[htb]
\noindent \centering{\includegraphics[width=0.5\textwidth]{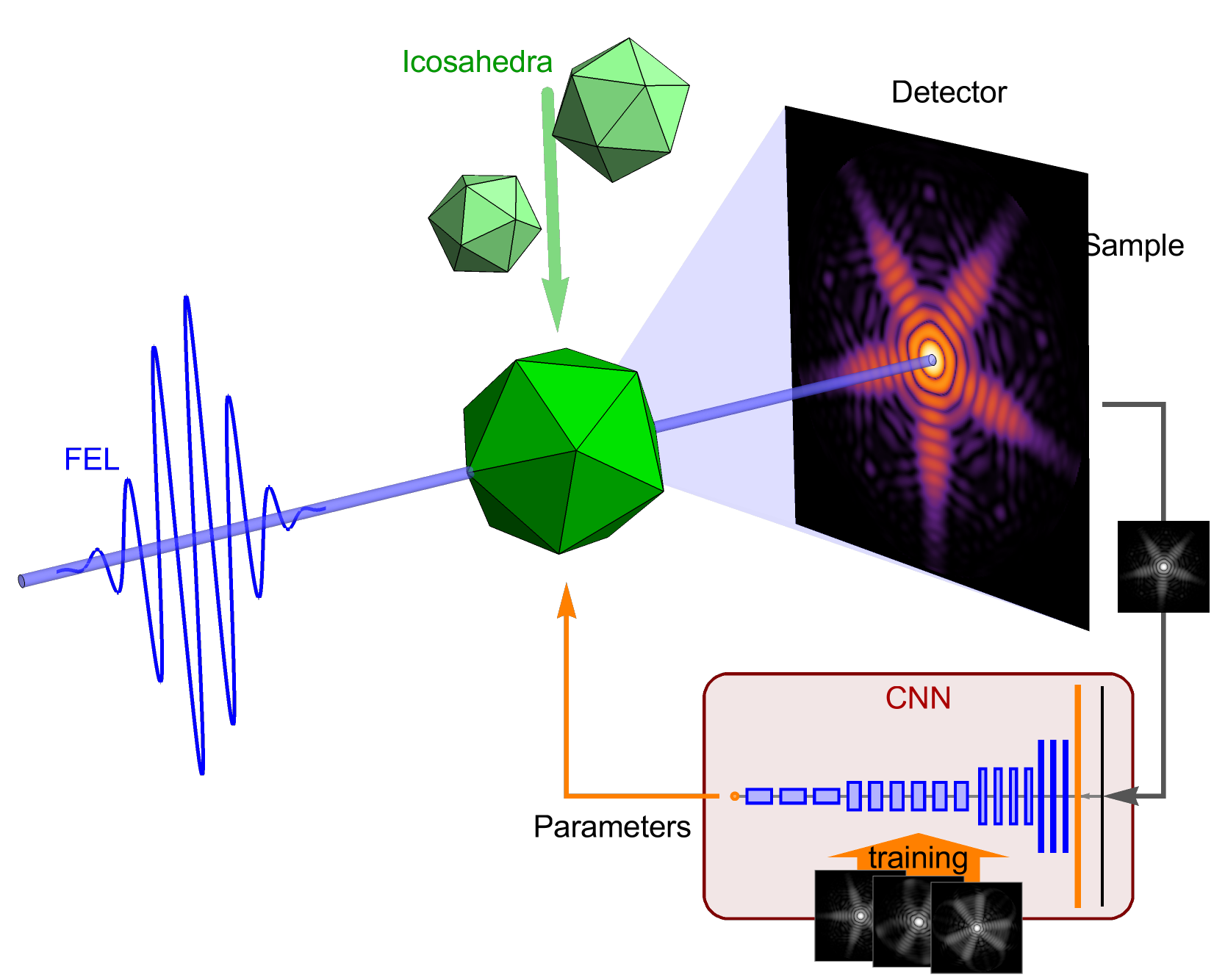}}
\caption{\textbf{Wide-angle scattering setup.} Nanoparticles of icosahedral
shape and varying size and orientation are interrogated by soft X-ray radiation 
from a free electron laser (FEL) \cite{Barke_2015}. The resulting wide-angle 
scattering pattern is simulated by employing an MSFT algorithm. The 
Convolutional Neural Network (CNN) computes a parameter representation of size 
and spatial orientation of the nanoparticle from the scattering 
image.}\label{fig:icoScattBeamline}
\end{figure}

The choice of icosahedra as test objects was motivated by their ubiquity in 
nature, ranging from viruses \cite{Seibert2011,Ekeberg_2015,Gorkhover2018,Ayyer_2019} 
to rare-gas \cite{Miehle1989} and metal clusters \cite{Barke_2015}. Focussing on 
the last example, which already constitutes a wide-angle scenario (see 
\fref{fig:icoScattBeamline}a), we compute scattering images of icosahedral 
silver clusters with a range of sizes and spatial orientations using an MSFT 
algorithm \cite{Langbehn_2018}, representing the training data. 
The employed generalized multi-sliced Fourier transform (MSFT) algorithm 
includes an effective treatment of absorption \cite{Barke_2015}. 

We numerically generated $\sim 25,000$ individual scattering images for clusters with a 
uniform size distribution ($30\,\mathrm{nm}\le R \le 160\,\mathrm{nm}$) and 
random orientations in the fundamental domain of the icosahedron, which 
represent perfect theoretical data. When representing spatial 
rotations by unit quaternions (see \ref{app:ico} for details), the 
fundamental domain of the icosahedron has the shape of a dodecahedron in 
imaginary space \cite{Frank_1986}, which is simply
connected and possesses a natural metric, unlike other lower dimensional representations such as 
Euler angles. Furthermore, any arbitrary rotation in the axis-angle representation may be projected into this 
domain by determining the distance to the closest quaternion associated to one 
of the symmetry rotations.

\begin{figure}[htb]
\noindent
\centering{\includegraphics[width=0.5\textwidth]{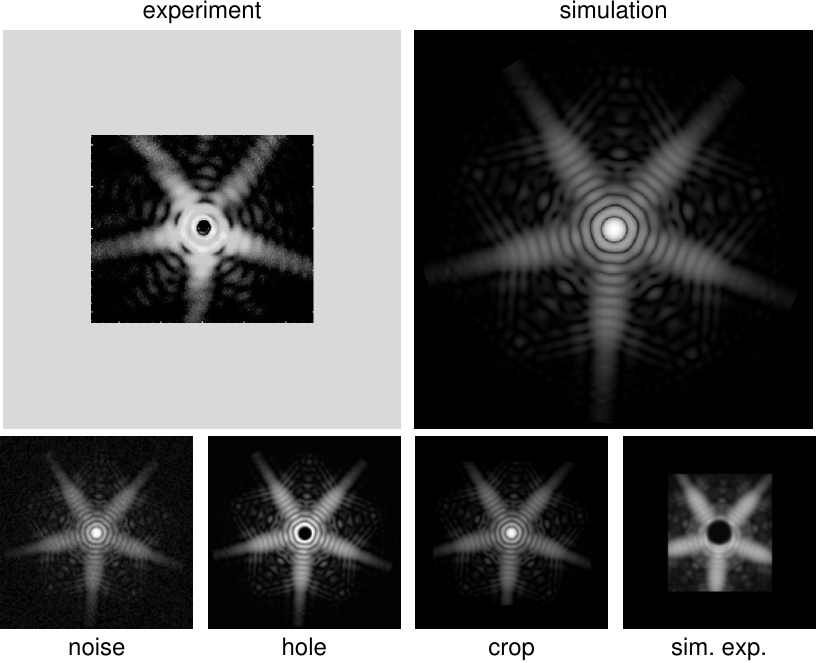}}
\caption{\textbf{Image Augmentations.} 
The ideal theoretical scattering images are augmented by image defects that 
account for experimental imperfections. They can be divided into quality 
defects such as noise or blur, and experiment-specific features such as the 
central hole and the limited size of the detector. We randomly combine all 
image effects, and in addition apply them in a well-defined order to 
generate images that closely resemble experimental data.
Experimental data taken from \cite{Barke_2015} (permitted by Creative Commons CC-BY 4.0 license (\url{http://creativecommons.org/licenses/by/4.0/)}).
}\label{fig:imageEffects}
\end{figure}

The ultimate goal is to analyze realistic scattering data that are obtained 
from experiments with various imperfections. Therefore, the neural network 
should not be trained solely using the ideal theoretical data, but also with 
appropriately augmented data \cite{Krizhevsky_2012,Dosovitskiy_2014,Perez_2017}. 
In that way, the network will be trained to focus on physically relevant features. 
Here, we augment our data by adding noise, blur, spatial offsets, a central hole, 
as well as blind spots and cropping of the images.
These augmentation features address common experimental imperfections 
associated with photon noise, limited detector resolution, source-point and 
beam-pointing jitter, transmission of the high-intensity primary beam, and 
detector segmentation and finite size (see \fref{fig:imageEffects}).
These augmentations (see \ref{app:aug} for details) increase the training set 11-fold.

\section{Network Design and Training}
Based upon the quaternion representation of rotations, we can find a 
unique, bijective parametrization of arbitrary icosahedra of uniform density by 
a vector with five scalar entries: the four components of the quaternions and 
the radius of the circumsphere. On the other hand, the associated scattering 
patterns can be understood as two-dimensional single-channel (or grayscale) 
images with a size of 128x128 pixels.
For the regression task of assigning a parameter vector to an image, we utilize 
the ResNet architecture of a convolutional neural network with 34 layers. 
This architecture was found to both offer the complexity needed for learning the
specified task while also keeping the number of free parameters as low as possible to 
counteract overfitting and minimizing training times.

\begin{figure}[htb]
\includegraphics[width=\textwidth]{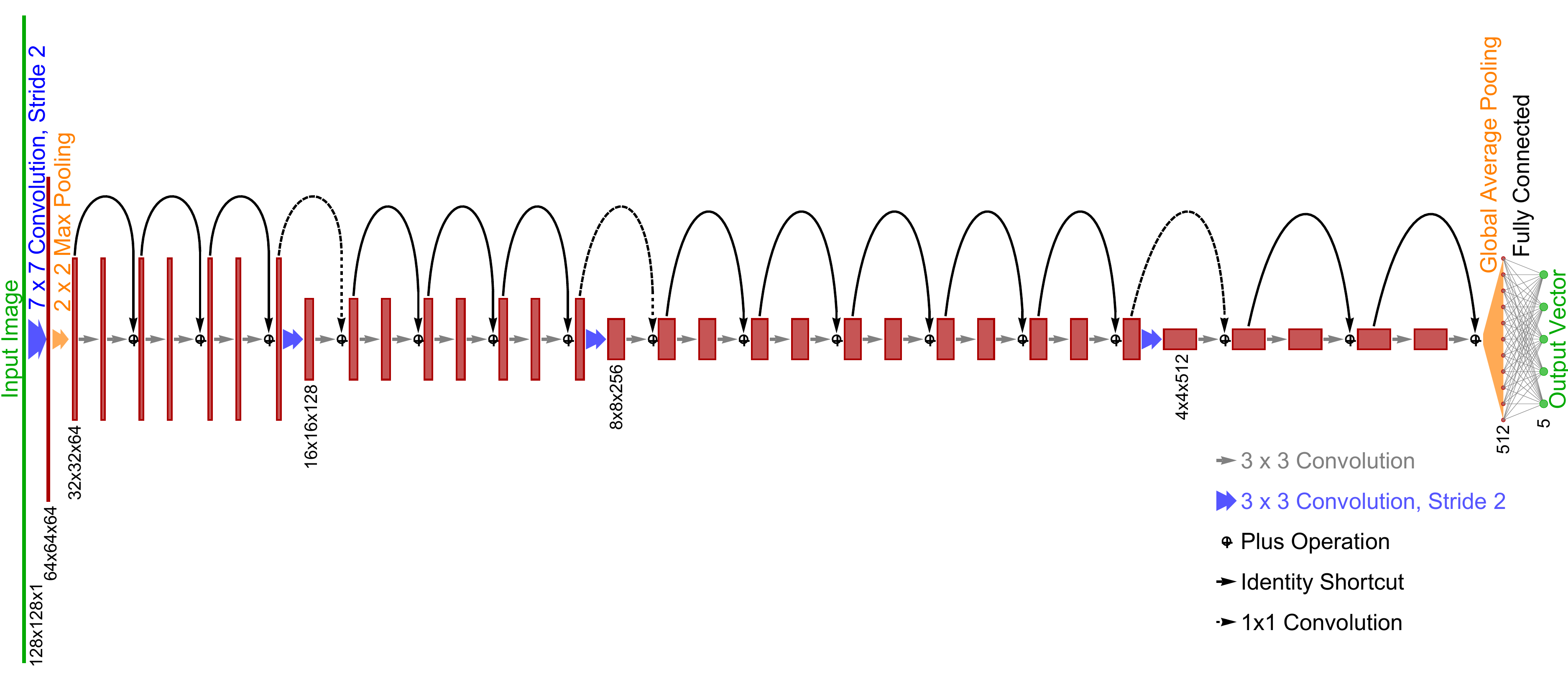}
\caption{\textbf{Network Design} The network architecture is based 
upon the ResNet scheme described in \cite{He_2016resNet}. The first two layers 
are used for an initial lateral dimension reduction, while all other convolution 
filters are encapsuled in residual blocks. Each block consists of two 
consecutive convolutional layers whose output is added to the initial input and,
by this, implement a residual calculation. The last residual block is fed into 
an average pooling operation compressing the tensor shape into a vector with 
512 entries from which the output five-vector is calculated by a fully 
connected layer. The five components of this output vector are assigned to the 
four components of a rotation quaternion and the radius, respectivley, and 
reprensent a full characterization of an icosahedron.}\label{fig:netScheme}
\end{figure}

The exact network structure is visualized in \Fref{fig:netScheme}. It is composed of an initial 7x7 convolution layer with stride 2 
and 64 filters, followed by a 2x2 max pooling for lateral dimension reductions, feeding into
four stacks of 3, 4, 6 and 4 consecutive residual blocks as defined in \cite{He_2016resNet} with 
64, 128, 256 and 512 filters each, respectively. The first convolution of each stack has stride 2
and consequently the corresponding identity shortcut is implemented by 1x1 convolutions. Behind the
final residual layer follows a global average pooling operation, reducing the tensor size to a 512-vector from
which the terminal 5-vector, composed of the four quaternion components and the radius, 
is computed by a fully connected layer. All activation functions are set to tanh.

Upon training, the network parameters were optimized to minimize the
mean-squared deviation of the predicted parameters compared to their target 
values. The training was performed on an Nvidia GTX 1060 consumer graphics 
card with the Wolfram language neural network framework, which was completed 
within approximately $4 \, \mathrm{h}$. During training, the generalization
capability is supervised by checking the prediction on the validation set after
each iteration on the training set.

\begin{figure}[htb]
\includegraphics[width=0.48\textwidth]{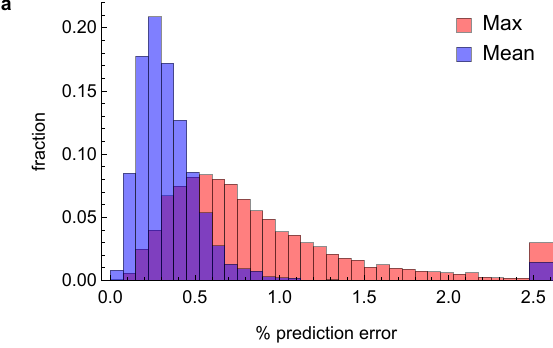}
\hspace*{\fill}
\includegraphics[width=0.48\textwidth]{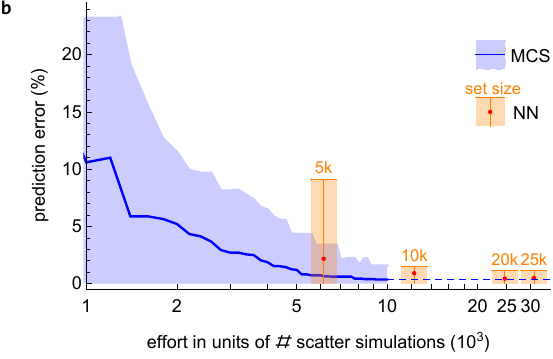}
\caption{\textbf{Performance Validation.} \textbf{a} The performance of the 
network is validated by computing the relative mean prediction error (blue bins).
The majority of the mean prediction errors is below $1\%$, with a minor quantile 
observing large errors that are mainly attributed 
to unphysical predictions. The maximal errors in each parameter (red bins) 
also remain mostly below $1\%$. 
\textbf{b} Evaluated on 30 random samples of the training set, the
Monte-Carlo-Simplex algorithm reached a median accuracy (blue line) of $0.37\%$
within 50 iterations. Each iteration step is estimated to require on average 
four scattering simulations, the horizontal axis denotes the number of 
scattering simulations during a single MCS run. The blue shaded area covers the 
$90\%$ quantile of the best-fit runs for each image, visualizing the error 
margin of the MCS method.
The performance of neural networks trained with subsets of the training set 
of different sizes are marked by red dots with shaded areas for the 
respective $90\%$ quantiles. The corresponding training times are also expressed
in units of scatter simulations. In order to achieve comparable median accuracy and 
error margin, the number of required scattering simulations for the training of 
a neural network corresponds to only a few MCS 
reconstructions.}\label{fig:paramHist}
\end{figure}

After training, the network's predictive capacity has been benchmarked on a
separate test set containing 5000 scattering patterns. The measure of interest
is the mean relative prediction error (normalized to the possible parameter range). 
In addition, we also specified the maximal relative prediction error out of the five 
parameters (for further information regarding error definitions see \ref{sec:errors}). 
These two errors (mean and maximum) are computed for each of the test patterns.
\Fref{fig:paramHist}a displays the resulting histograms for the mean (blue 
bins) and the maximum prediction error (red bins). The reconstruction of the 
relevant physical parameters is highly accurate, with prediction errors well 
below $1\%$.

For comparison with established forward fitting methods, we also reconstructed 
30 images of the test set with a state-of-the-art Monte Carlo simplex procedure, 
as used in \cite{Langbehn_2018}. For each image, the reconstruction started 
from 50 random initial points in parameter space, and 50 simplex iteration steps 
have been taken. The convergence of the reconstruction error as function of 
required image calculations can be seen in \fref{fig:paramHist}b, where we 
estimated that, on average, four scattering patterns need to be calculated in 
each iteration step. The solid line marks the median best approximation, while the 
shaded area marks the $90\%$ quantile. In comparison, the red bars denote the 
same measures for individual neural networks trained on different sized
portions of the complete training set. It can be seen that both methods eventually 
achieve the same level of accuracy. The optimized scattering code requires 
$\sim 2.5\,\mathrm{s}$ per image on a hexa-core Intel Xeon E5, resulting in a 
generation time of $\sim 17\,\mathrm{h}$ for the complete training set. 
The subsequent training of the network on the complete data set takes additional 
$\sim 4\,\mathrm{h}$, which result in a total of $\sim 21\,\mathrm{h}$ or the 
equivalent of 31k scattering image calculations to yield the ready-to-use neural 
network. After successful training, the evaluation of a single image takes only 
$5\,\mathrm{ms}$, which is a negligible numerical effort compared to that 
required during forward fitting. Hence, already for a small number of to be 
reconstructed images, the computational overhead for training data set 
generation and actual network training can be compensated by the exceptional 
reconstruction speed whilst still providing reconstruction results of comparable 
accuracy (see \fref{fig:paramHist}b).

\begin{figure}[htb]
\centering{\includegraphics[width=0.6\columnwidth]{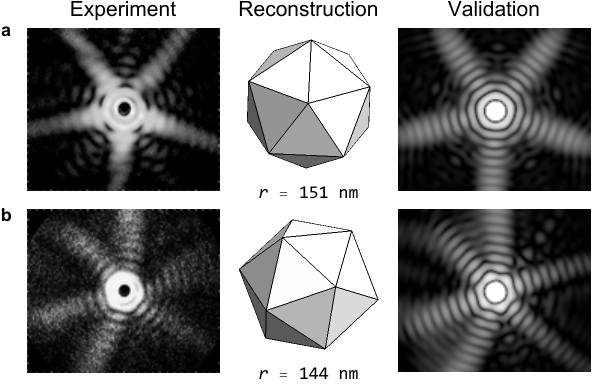}}
\caption{\textbf{Reconstruction from experimental data.} Experimental data 
from \cite{Barke_2015} (left column, permitted by Creative Commons CC-BY 
4.0 license (\url{http://creativecommons.org/licenses/by/4.0/})), are evaluated 
by the neural network. The reconstructed spatial orientation in the laser 
propagation direction is shown in the middle column. The reconstructed radii 
are very close to those given in \cite{Barke_2015}. The theoretical 
scattering patterns associated with these reconstructions reproduce the 
experimental images very well, including low-intensity features (right 
column). The intensity of the theoretical patterns is clipped at a maximum intensity
in order to mimic the nonlinear response of the detector.
}\label{fig:expPlot}
\end{figure}

We demonstrate the network's ability in recognizing structures in imperfect
experimental images by applying it to data taken from \cite{Barke_2015},
where two icosahedral clusters have been identified among the images (left 
column in \fref{fig:expPlot}). The reconstructed size and spatial 
orientation (central column in \fref{fig:expPlot}) are validated to 
reproduce the experimental scattering images (right column in 
\fref{fig:expPlot}). Our results match the 
reconstructed data published in \cite{Barke_2015}, with the exception of 
one of the radii which we attribute to the reduced visibility of the 
radial fringes which complicates an accurate radius determination with any method.

\section{Summary}
We have shown that, using a deep-learning technique based on augmented 
theoretical scattering data, neural networks enable the accurate and fast 
reconstruction of wide-angle scattering images of individual icosahedral 
nanostructures. Our results demonstrate that a network, which has only 
been trained on theoretical data, can be employed for the analysis 
of experimental scattering data, with image processing times on the millisecond 
time scale. The neural network reaches the same level of accuracy as 
established forward fitting methods based on Monte Carlo Simplex algorithms.
Although the reconstruction of a single image using the neural network is 
orders of magnitude faster than the direct optimization, the generation of the 
training data and subsequent training of the network requires a substantial 
constant overhead. However, the reconstruction speed of the network compensates 
the extra effort after only a few scattering images.

Motivated by the performance of this method, we anticipate that a 
generalization to a wide range of particle morphologies will be feasible.
Combined with pre-selection algorithms as utilized in \cite{Langbehn_2018},
this may evolve into a classification tool for archimedean bodies.
The envisaged combination of identification of arbitrary three-dimensional 
shapes with short processing times is anticipated to represent the enabling 
technology for a fully automated analysis of scattering data and real-time 
reconstruction of ultrafast nanoscale dynamics probed at the next generation of 
X-ray light sources with high repetition rate --- with major implications for a 
broad range of physical, chemical and biological applications. 

\ack
T.S. gratefully acknowledges a scholarship from ``Evangelisches Studienwerk 
Villigst''. S.S. acknowledges financial support from Deutsche 
Forschungsgemeinschaft (DFG) via the SPP 1929 ``Giant interactions in 
Rydberg systems''. T.F. acknowledges financial support from the DFG via the 
Heisenberg program (No. 398382624) and the BMBF (project 05K16HRB). This work 
was partially supported by the NEISS project of the European Social Fund (ESF) 
(reference ESF/14-BM-A55-0007/19) and the Ministry of Education, Science and 
Culture of Mecklenburg-Vorpommern, Germany.  The authors thank Dr. Katharina 
Sander for sharing her insights into the MSFT technique and
Dr. Ingo Barke for providing information on the experimental data aquisition.

\section*{Data Availability}
The data that support the findings of this study are available from the corresponding author upon reasonable request.

\appendix
\setcounter{section}{0}
\section{Icosahedral Symmetry}\label{app:ico}
The icosahedron is one of the five platonic solids and is spanned by 20 
equilateral triangle faces, intersecting with 30 edges and twelve corners. 
It possesses three-fold rotation symmetry axes $C_3$ about the center-of-mass 
of each triangle, two-fold axes $C_2$ about the center of each edge 
and five-fold axes about each corner, which together form the icosahedral 
rotation group $I$. The 60 symmetry rotations imply that any rotation 
of a body with icosahedral symmetry is 60-fold degenerate. 
Hence, the mapping of three-dimensional rotation representations, such as 
Euler-angle or axis-angle representations, to icosahedral orientations are not 
unique, but have to be constrained in their parameter range. The fundamental 
domain of rotations has an exceptionally simple form in quaternion 
representation of rotations, where it forms a dodecahedron in imaginary space 
\cite{Frank_1986}. 

Quaternions $\mathbb Q$ are the four-dimensional extension of the complex 
numbers with three imaginary units $\ii, \jj$ and $\kk$ fulfilling the
relations $\ii^2 = \jj^2 = \kk^2 =\ii \jj \kk = -1$ and $\ii \jj = - \jj \ii$. 
With real coefficients $q_i$, any quaternion may be written as 
$q = q_0 + \ii q_1 +\jj q_2 + \kk q_3$.
Imaginary quaternions ($q_0 \equiv 0$) are isomorphic to the space 
$\mathbb R^3$, implying that all vectors $\mathbf a = (a_1, a_2, a_3)$ can be 
represented by quaternions as $q_a = \ii a_1 + \jj a_2 + \kk a_3$. The sum of 
two vectors then translates into the sum of two quaternions, whereas the 
quaternion product contains both the scalar product of two cartesian vectors 
(in its real part) and their cross product (in the imaginary part). The 
rotation by an angle $\alpha$ of any vector $\mathbf{a}$ about a unit vector 
$\mathbf{n}$ can thus be expressed by the product of the quaternion $q_a$ with 
the unit quaternion 
$q_\mathrm{rot} = \cos(\alpha/2) + \sin(\alpha/2)\,(n_x \ii + n_y \jj + n_z 
\kk)$. 
Hence, any rotation can be projected into the fundamental domain by applying 
all inverse symmetry rotations and selecting the one yielding the smallest 
rotation angle. For the training of a neural network, the quaternion 
representation has the additional advantage of providing a useful metric for 
the distance between rotations. 

\section{Dataset Generation}	
The scattering patterns used for training are created by using the MSFT
algorithm described in ~\cite{Barke_2015}. In accordance with the
experiment described therein, we simulate the scattering of ultra-short
XUV pulses with wavelength $\lambda = 13.5 \, \mathrm{nm}$ and femto-second
duration on nano-sized silver clusters. The material parameters are assumed
to be equal to bulk silver, with absorption length
$a_\mathrm{abs} = 12.5 \, \mathrm{nm}$. For the calculations, the electron density
of the cluster is discretized on a cuboid grid, chosen to contain a depth of
192 pixels. The outgoing scattered field is determined by the phase-coherent
summation of the scattered field of each slice, which can be obtained by
Fourier transformation. Before transformation, each slice is zero-padded to a
width of 512 pixels. The computed scattered field is then reduced to an logarithmic 
intensity profile of $128\times 128$ pixels with random background noise, which 
is stored as a grayscale image. The rotation quaternions are sampled
uniformly from the fundamental domain, while the size of the clusters range from
$30$ to $160 \, \mathrm{nm}$. With this procedure, a dataset of $25\,058$ images has
been generated, one fifth of which has been reserved for validation during training. 
Another set of $5000$ is generated for final testing indepentently from the training process.

\section{Error Measures}\label{sec:errors}
The five parameters of our icosahedron representation reconstructed by the neural network 
cover very different parameter ranges. For training purposes, both the parameters of the radius
and the real part of the rotation quaternion are linearly scaled to the interval $[0,1]$. However, 
when testing the prediction quality of the network it is more useful to have a uniform error measure,
that gives the same weight to each parameter. Hence, all error measures within this work are calculated from the relative errors of each prediction parameter, obtained by normalization to the allowed parameter range. All four
components of the rotation quaternion are normalized to the maximal span of the fundamental domain in each
direction. The deviation of the radius is normalized to the range defined in the previous section.
These five individual relative errors, normalized to the parameter range allowed in our model, now weigth each
parameter equally. From these, further error measures can be defined. More precisely, the mean error, taking the 
mean values of all five relative errors, and the maximum error, picking the largest of all five relative errors.

\section{Image Augmentation}\label{app:aug}
Prior to training the neural network, image augmentation is applied to the 
dataset. The augmentation is performed by applying eleven different filters to 
each ideal scattering pattern, and randomly adding the newly generated images 
to the training set. These filters can be divided into five groups: 
trivial, noise, blur, cropping and successive application. The 
trivial filter is the identity mapping, leaving the image unchanged. Noise is 
applied both with uniform distribution with a randomly chosen intensity upto 
half the maximum signal, changing every pixel by a random margin 
as well as salt-and-pepper statistics, where random pixels are set to either 
minimal or maximal signal. Blurring is performed by convoluting with a 
Gaussian kernel with randomly chosen radius of upto five pixels, and by jitter 
distortion. Cropping filters delete different 
parts of the image, mainly to account for the characteristics of real 
detectors. Images are either center-cropped for limited detector size, a central 
hole of random radius is deleted to simulate the shadow of a beam dump, 
images are shifted or uneven detector sensitivity is simulated by attenuating 
parts of the image. Finally, we both randomly combine all image effects, and in 
addition apply them in a well-defined order so as to generate images 
that closely resemble experimental results (see \fref{fig:imageEffects}).


\section*{References}
\bibliographystyle{iopart-num}
\bibliography{clusterNNBib}

\providecommand{\newblock}{}
\begin{thebibliography}{10}
\expandafter\ifx\csname url\endcsname\relax
  \def\url#1{{\tt #1}}\fi
\expandafter\ifx\csname urlprefix\endcsname\relax\def\urlprefix{URL }\fi
\providecommand{\eprint}[2][]{\url{#2}}

\bibitem{Chapman2006}
Chapman H~N, Barty A, Bogan M~J, Boutet S, Frank M, Hau-Riege S~P, Marchesini
  S, Woods B~W, Bajt S, Benner W~H {\em et~al.\/} 2006 {\em Nature Physics\/}
  {\bf 2} 839

\bibitem{Gaffney2007}
Gaffney K and Chapman H 2007 {\em Science\/} {\bf 316} 1444--1448

\bibitem{Seibert2011}
Seibert M~M, Ekeberg T, Maia F~R, Svenda M, Andreasson J, J{\"o}nsson O,
  Odi{\'c} D, Iwan B, Rocker A, Westphal D {\em et~al.\/} 2011 {\em Nature\/}
  {\bf 470} 78

\bibitem{Gomez2014}
Gomez L~F, Ferguson K~R, Cryan J~P, Bacellar C, Tanyag R~M~P, Jones C, Schorb
  S, Anielski D, Belkacem A, Bernando C {\em et~al.\/} 2014 {\em Science\/}
  {\bf 345} 906--909

\bibitem{Rupp_2017}
Rupp D, Monserud N, Langbehn B, Sauppe M, Zimmermann J, Ovcharenko Y,
  M{\"o}ller T, Frassetto F, Poletto L, Trabattoni A {\em et~al.\/} 2017 {\em
  Nature communications\/} {\bf 8} 493

\bibitem{Langbehn_2018}
Langbehn B, Sander K, Ovcharenko Y, Peltz C, Clark A, Coreno M, Cucini R,
  Drabbels M, Finetti P, Di~Fraia M {\em et~al.\/} 2018 {\em Physical review
  letters\/} {\bf 121} 255301

\bibitem{Rupp2012}
Rupp D, Adolph M, Gorkhover T, Schorb S, Wolter D, Hartmann R, Kimmel N, Reich
  C, Feigl T, De~Castro A {\em et~al.\/} 2012 {\em New Journal of Physics\/}
  {\bf 14} 055016

\bibitem{Barke_2015}
Barke I, Hartmut H, Rupp D, Fl{\"u}ckiger L, Sauppe M, Adolph M, Schorb S,
  Bostedt C, Treusch R, Peltz C, Bartling S, Fennel T, Meiwes-Broes K~H and
  M{\"o}ller T 2015 {\em Nature communications\/} {\bf 6} 6187

\bibitem{Ekeberg_2015}
Ekeberg T, Svenda M, Abergel C, Maia F~R, Seltzer V, Claverie J~M, Hantke M,
  J{\"o}nsson O, Nettelblad C, Van Der~Schot G {\em et~al.\/} 2015 {\em
  Physical review letters\/} {\bf 114} 098102

\bibitem{Ayyer_2019}
Ayyer K, Morgan A~J, Aquila A, DeMirci H, Hogue B~G, Kirian R~A, Xavier P~L,
  Yoon C~H, Chapman H~N and Barty A 2019 {\em Optics Express\/} {\bf 27}
  37816--37833

\bibitem{Raines2010}
Raines K~S, Salha S, Sandberg R~L, Jiang H, Rodr{\'\i}guez J~A, Fahimian B~P,
  Kapteyn H~C, Du J and Miao J 2010 {\em Nature\/} {\bf 463} 214

\bibitem{Gessner2019}
Gessner O and Vilesov A 2019 {\em Annual Review of Physical Chemistry\/} {\bf
  70}

\bibitem{hinton2006}
Hinton G~E and Salakhutdinov R~R 2006 {\em Science\/} {\bf 313} 504--507

\bibitem{lecun2015}
LeCun Y, Bengio Y and Hinton G 2015 {\em Nature\/} {\bf 521} 436

\bibitem{carleo2019review}
Carleo G, Cirac I, Cranmer K, Daudet L, Schuld M, Tishby N, Vogt-Maranto L and
  Zdeborov{\'a} L 2019 {\em arXiv preprint arXiv:1903.10563\/}

\bibitem{wang2016}
Wang L 2016 {\em Physical Review B\/} {\bf 94} 195105

\bibitem{carrasquilla2017}
Carrasquilla J and Melko R~G 2017 {\em Nature Physics\/} {\bf 13} 431

\bibitem{van2017}
Van~Nieuwenburg E~P, Liu Y~H and Huber S~D 2017 {\em Nature Physics\/} {\bf 13}
  435

\bibitem{venderley2018}
Venderley J, Khemani V and Kim E~A 2018 {\em Physical Review Letters\/} {\bf
  120} 257204

\bibitem{carleo2017}
Carleo G and Troyer M 2017 {\em Science\/} {\bf 355} 602--606

\bibitem{torlai2018}
Torlai G, Mazzola G, Carrasquilla J, Troyer M, Melko R and Carleo G 2018 {\em
  Nature Physics\/} {\bf 14} 447

\bibitem{xin2018}
Xin T, Lu S, Cao N, Anikeeva G, Lu D, Li J, Long G and Zeng B 2019 {\em npj
  Quantum Information\/} {\bf 5} 1--8

\bibitem{Cherukara_2018}
Cherukara M~J, Nashed Y~S and Harder R~J 2018 {\em Scientific reports\/} {\bf
  8} 1--8

\bibitem{laanait2019}
Laanait N, He Q and Borisevich A~Y 2019 {\em arXiv preprint arXiv:1902.06876\/}

\bibitem{laanait2019exascale}
Laanait N, Romero J, Yin J, Young M~T, Treichler S, Starchenko V, Borisevich A,
  Sergeev A and Matheson M 2019 {\em arXiv preprint arXiv:1909.11150\/}

\bibitem{Zimmermann2019}
Zimmermann J, Langbehn B, Cucini R, Di~Fraia M, Finetti P, LaForge A~C,
  Nishiyama T, Ovcharenko Y, Piseri P, Plekan O {\em et~al.\/} 2019 {\em
  Physical Review E\/} {\bf 99} 063309

\bibitem{Gorkhover2018}
Gorkhover T, Ulmer A, Ferguson K, Bucher M, Maia F~R, Bielecki J, Ekeberg T,
  Hantke M~F, Daurer B~J, Nettelblad C {\em et~al.\/} 2018 {\em Nature
  Photonics\/} {\bf 12} 150

\bibitem{Miehle1989}
Miehle W, Kandler O, Leisner T and Echt O 1989 {\em The Journal of chemical
  physics\/} {\bf 91} 5940--5952

\bibitem{Frank_1986}
Frank F~C 1986 {\em Le Journal de Physique Colloques\/} {\bf 47} C3--165

\bibitem{Krizhevsky_2012}
Krizhevsky A, Sutskever I and Hinton G~E 2012 Imagenet classification with deep
  convolutional neural networks {\em Advances in neural information processing
  systems\/} pp 1097--1105

\bibitem{Dosovitskiy_2014}
Dosovitskiy A, Springenberg J~T, Riedmiller M and Brox T 2014 Discriminative
  unsupervised feature learning with convolutional neural networks {\em
  Advances in Neural Information Processing Systems\/} pp 766--774

\bibitem{Perez_2017}
Perez L and Wang J 2017 {\em arXiv preprint arXiv:1712.04621\/}

\bibitem{He_2016resNet}
He K, Zhang X, Ren S and Sun J 2016 Deep residual learning for image
  recognition {\em Proceedings of the IEEE conference on computer vision and
  pattern recognition\/} pp 770--778

\end{thebibliography}

\end{document}